\documentclass[superscriptaddress]{PoS}
\usepackage{url}
\usepackage{amssymb}
\usepackage{multicol}

\graphicspath{{figures/}}

\title{Temporal analysis of an unprecedented data set for the $\gamma$-ray blazar 1ES 1215+303:\\ {\bf\it Fermi}-LAT and VERITAS light curves spanning ten years}

\ShortTitle{Unprecedented temporal analysis of the $\gamma$-ray blazar 1ES 1215+303}

\author{\speaker{Janeth Valverde}$^{\dagger}$, Deirdre Horan$^{\dagger}$, Giuliana Noto$^{\ddagger}$, Reshmi Mukherjee$^{\ddagger}$, Denis Bernard$^{\dagger}$ \\
	 $^{\dagger}$LLR/Ecole Polytechnique, Palaiseau, France.\\$^{\ddagger}$Barnard College, Columbia University, New York, NY, United States.\\
        E-mail: \email{valverde@llr.in2p3.fr}}
\author{on behalf of the {\bf\it Fermi}-LAT \& VERITAS Collaborations}

\abstract{We present here the results of the analysis of the $\gamma$-ray blazar, 1ES 1215+303, over a 10-year period, from 2008 to 2017, measured at high energies (HE; 200 MeV $< E <$ 100 GeV) by the 
    Fermi Large Area Telescope (LAT) and at very high energies (VHE; $E >$ 100 GeV) by {\it Fermi}-LAT and VERITAS. This is the longest temporal study of this high-frequency-peaked BL Lac object (HBL) at 
    $\gamma$-ray energies to date. The spectrum follows a log parabola over this time period, and its HE and VHE spectra are well-connected. Its flux is sufficiently strong at HE to allow us to bin the 
    {\it Fermi}-LAT data in 3-day intervals, enabling us to investigate the temporal evolution of the flux in unprecedented detail. Several flaring episodes were detected and evidence for an overall trend 
    of increasing flux over the span of the 10 years was observed. These light curves, in addition to the spectra, are presented. This unique data set will help us to advance our understanding 
    of the underlying physical processes in blazar jets.}

\FullConference{7th Fermi Symposium 2017\\
		15-20 October 2017\\
		Garmisch-Partenkirchen, Germany}

\begin{document}

\section{Introduction}
The largest population of sources detected by the current generation of space-based and ground-based $\gamma$-ray telescopes are blazars. 
They comprise more than 50\% of the associated sources in the Third Fermi catalogue (3FGL) \cite{Acero}, more than 70\% of the Third Catalog of Hard {\it Fermi}-LAT Sources \cite{3FHL}, and more than 30\% of the 
objects detected at TeV energies\footnote{\url{http://tevcat2.uchicago.edu/}}. Blazars are a subcategory of active galactic nuclei, very powerful systems believed to be powered by a 
central supermassive black hole, in which the relativistic jet is approximately coaligned with the line of sight of the Earth \cite{beck}. BL Lac 
objects are a further subclass of blazars, which usually do not have significant emission or absorption features in their optical spectra \cite{giebels}. The study of high-flux
variability of blazars allows us to put constraints on the size and location of the emitting region, while at the same time enabling us
to probe the particle acceleration mechanisms in the blazar jets. Detailed studies across the entire $\gamma$-ray range allow the study 
the extragalactic background light (EBL, \cite{hauser}) due to the TeV photon absorption by pair production, which can be used to estimate the 
density of the EBL \cite{eblblazars}. \\
1ES 1215+303 (B2 1215+30, ON 325) is a blazar located at R. A.$_{(J2000)}=12^h17^m48.5^s$ and Dec.$_{(J2000)}=+30^\circ07'00''6$. It was first reported
in the 1970 B2 408 MHz Bologna Northern Cross telescope radio catalogue \cite{Colla} (hence the name B2),  and has been observed at other wavelengths since 
then, including its first detection at VHE on 2011-01-[02-05] by the MAGIC collaboration (ATel \#3100, \cite{aleksic}). It has been 
classified as a high-frequency peaked \cite{ack2011} BL Lac \cite{browne} with a power-law spectrum, and its redshift was recently confirmed to be $z =$ 0.131, by optical spectroscopy \cite{paiano}. \\
There have been two separate detections of flaring activities from 1ES 1215+303 at VHE reported by VERITAS \cite{muk}, and another two at HE, detected with the 
{\it Fermi}-LAT \cite{abdo}\cite{muk}. The first HE flare for this source occurred on 2008-10-[10-15] with a reported flux of F$_{(E>300\mbox{\footnotesize MeV})}\sim 15\times 10^{-8}$cm$^{-2}$s$^{-1}$ 
in a weekly light curve as part of a broader 
study of variability that included another 105 blazars \cite{abdo}. The second HE flare occurred with a coincident counterpart at VHE on 2014-02-08 and was described in a detailed 
publication \cite{muk}, which also reported the 2013-02-07 VHE flare. Moreover, 1ES 1215+303 
was the subject of a long-term multi-wavelength study \cite{aliue} that covered the time range from 2008 to 2012 with a  8.9 $\sigma$ significance 
for energies greater than 100 GeV. In this contribution we report our preliminary results on the long-term variability and characteristics of 1ES 1215+303 
in its different states using data collected with the {\it Fermi}-LAT and VERITAS from 2008 to 2017.

\section{Instruments}
\subsection{VERITAS}
The Very Energetic Radiation Imaging Telescope Array System (VERITAS)
is an array of four imaging atmospheric Cherenkov telescopes located
at the Fred Lawrence Whipple Observatory in southern Arizona,
U.S.A. (31$^{\circ}$ 40'N, 110$^{\circ}$ 57'W, 1.3km a.s.l.). Each of the four
telescopes comprises a 12-m diameter mirror, of the Davies Cotton
design, and an imaging camera with 499 high-quantum-efficiency
photomultiplier tubes. The four telescopes operate in stereo mode to cover a field of 
view of approximately 3.5$^{\circ}$ with an angular resolution of
0.08$^{\circ}$ at 1 TeV (68\% containment radius). The energy range of VERITAS
goes from 85 GeV to above 30 TeV (spectral reconstruction possible
from 100 GeV) with an energy resolution of 15-25\% and an effective
collection area for a 1 TeV photon on the order of 10$^5$m$^2$. A source
with a flux of 1\% that of the Crab Nebula, the standard candle at
these energies, is detectable at 5 standard deviations above
 background in approximately 25 hours \cite{holder}.
 The data were analysed using the two
standard VERITAS analysis software packages \cite{acciari}, and produced
results that were in excellent agreement with each other.

\subsection{{\it Fermi}-LAT}
The Large Area Telescope (LAT), on board
the {\it Fermi Gamma-ray Space Telescope}, is a pair-conversion detector that covers the energy range from 20 MeV
to more than 500 GeV with a field of view of
approximately 2.4 sr \cite{latpaper}. The main observation mode of the {\it Fermi}-LAT is survey mode during which the LAT scans the entire sky every 3 hrs.
We analyzed the {\it Fermi}-LAT data for
1ES 1215+303 from 2008-08-04, the start of the mission, up until to 2017-09-04. The data analysis was
performed using the Fermi Science Tools package,
version v10r0p5, and the latest P8R2\_SOURCE\_V6
instrument response functions. We have considered a
maximum zenith angle of 90$^{\circ}$ in order to reduce
contributions from the Earth limb; and photons with
energies greater than 100 MeV. For each data sample, we considered photons in a 10$^{\circ}$ region of interest centered at the position of 1ES 1215+303;
and contributions from sources from the 3FGL within a 16.2$^{\circ}$ region were included. None of the residual maps obtained in our analysis showed evidence 
for any additional excess. For the modeling of 1ES 1215+303, we considered both a power-law (PL) and a log-parabola (LP) spectral model (the second simplest model used in the 3FGL).
\vspace{-4mm}
 \begin{figure}
  \centering\includegraphics[width=0.28\textwidth]{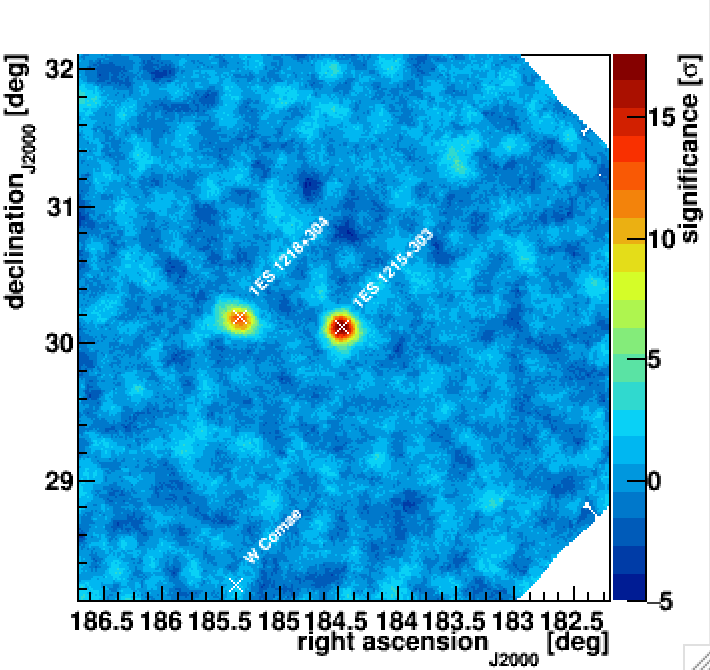}\hspace{1.5mm}\includegraphics[width=0.36\textwidth]{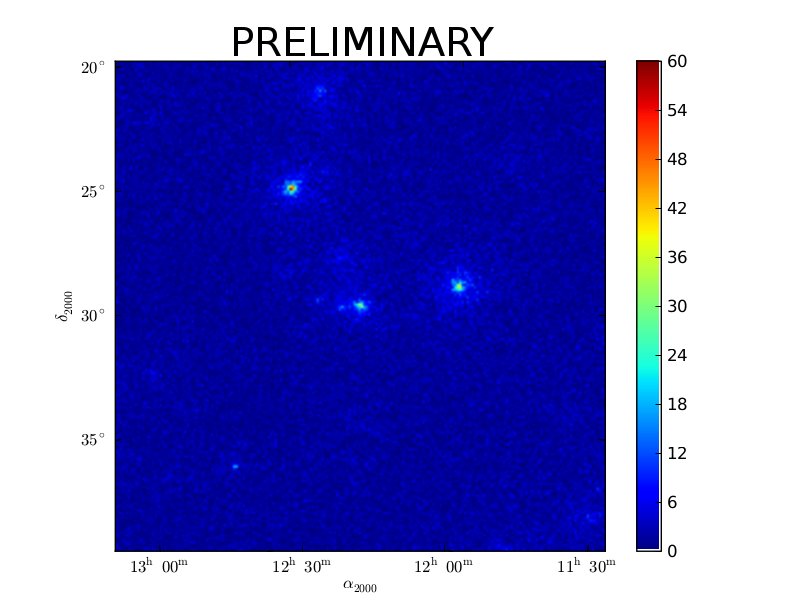}
  \caption{VERITAS (left) and LAT (right) 2015 1ES 1215+303 Skymaps.}
     \end{figure}  
     \vspace{-5.5mm}
\section{Preliminary results}
\begin{figure}\includegraphics[width=0.99\textwidth]{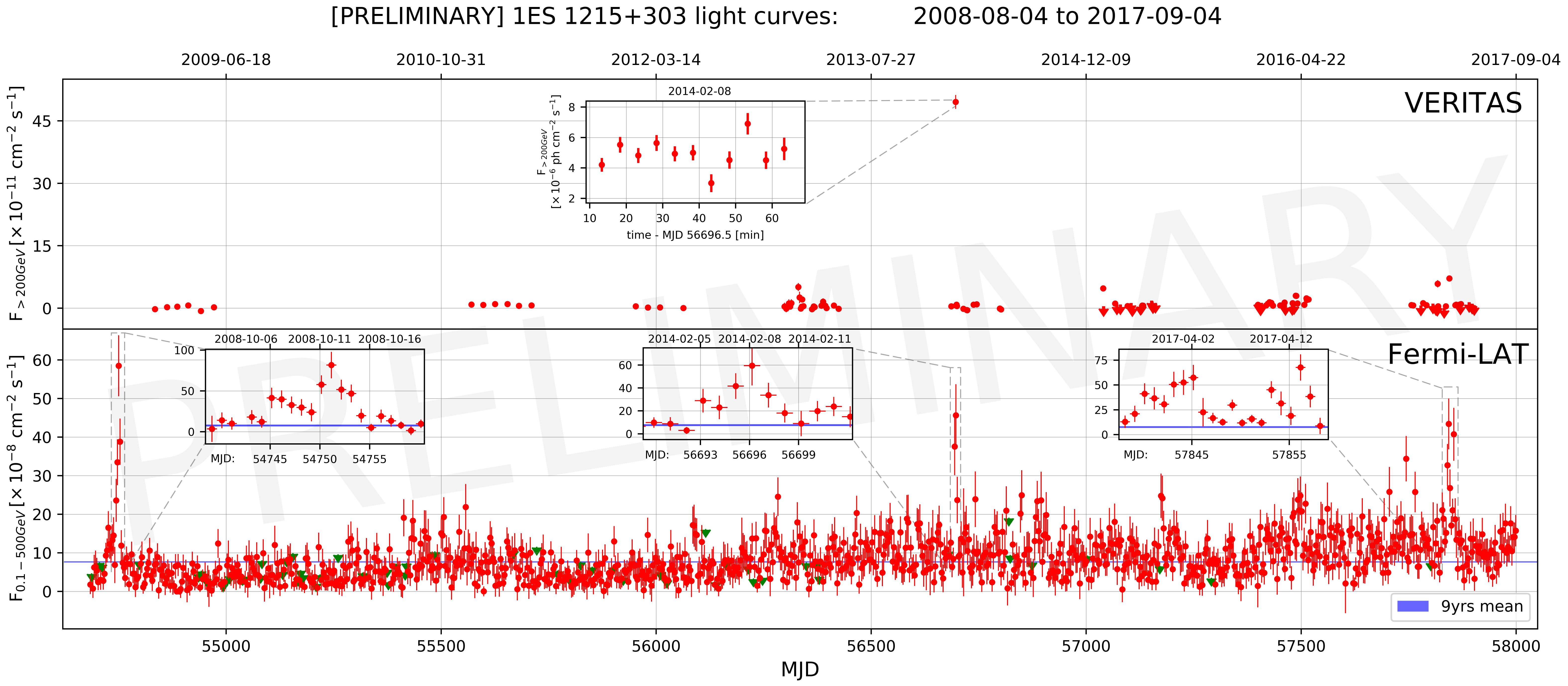}
      \caption{Long-term light curves: VERITAS 1-day data points on the top and 3-day LAT data on the bottom.}
       \label{fig:ltlc}  
\vspace{2mm}       
    \begin{minipage}{0.495\textwidth}
  \includegraphics[width=0.99\textwidth]{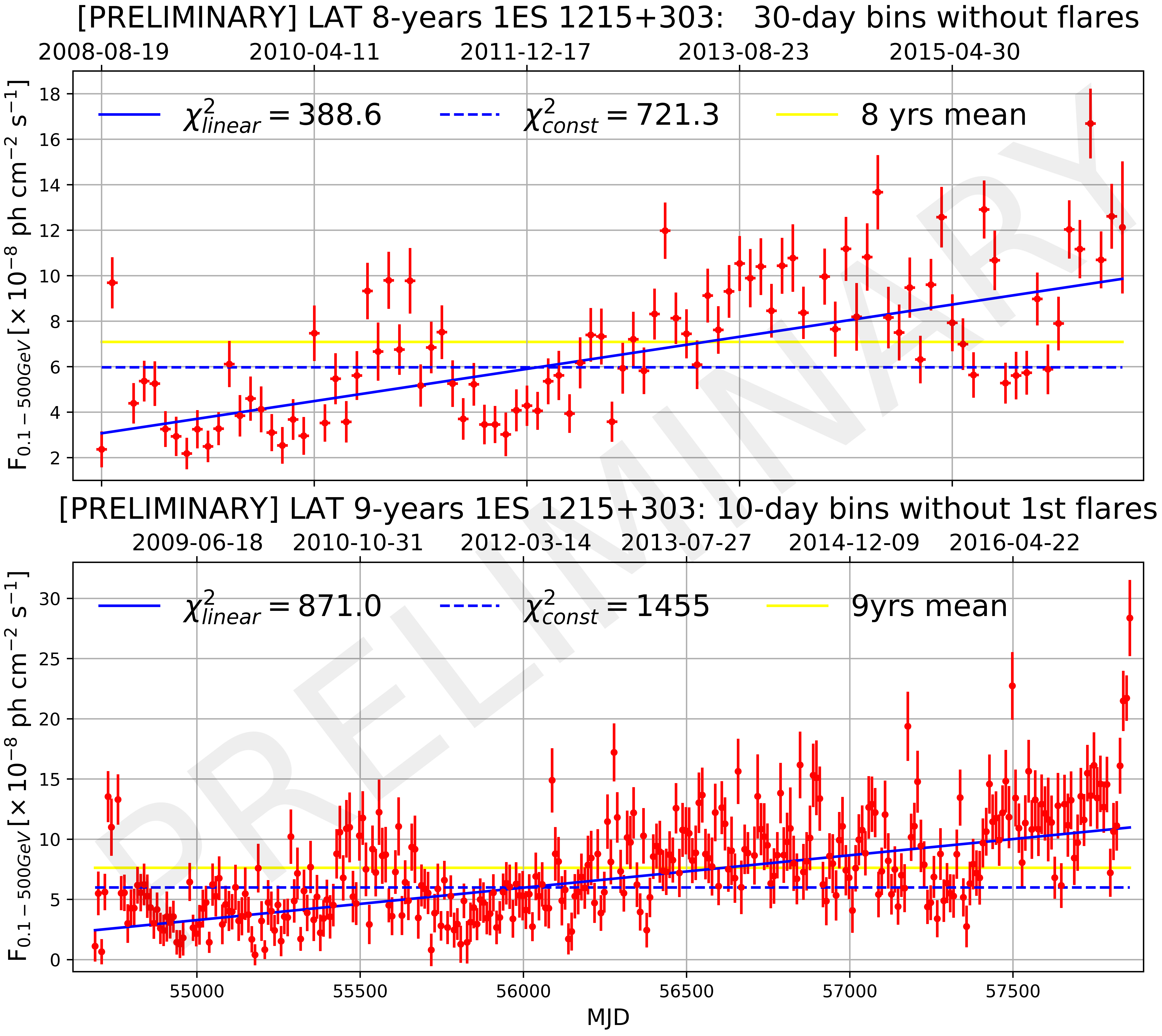}  
   \caption{{\textit Top}: 30-day binned light curve. {\textit Bottom}: 10-day binned light curve. Both of them exclude the brightest flares. 
   We observe a preference for a linear increase of the flux over a constant average in both data sets.}
      \label{fig:1030}
      \end{minipage}
\hfill
    \begin{minipage}{0.495\textwidth}
     \includegraphics[width=0.5\textwidth]{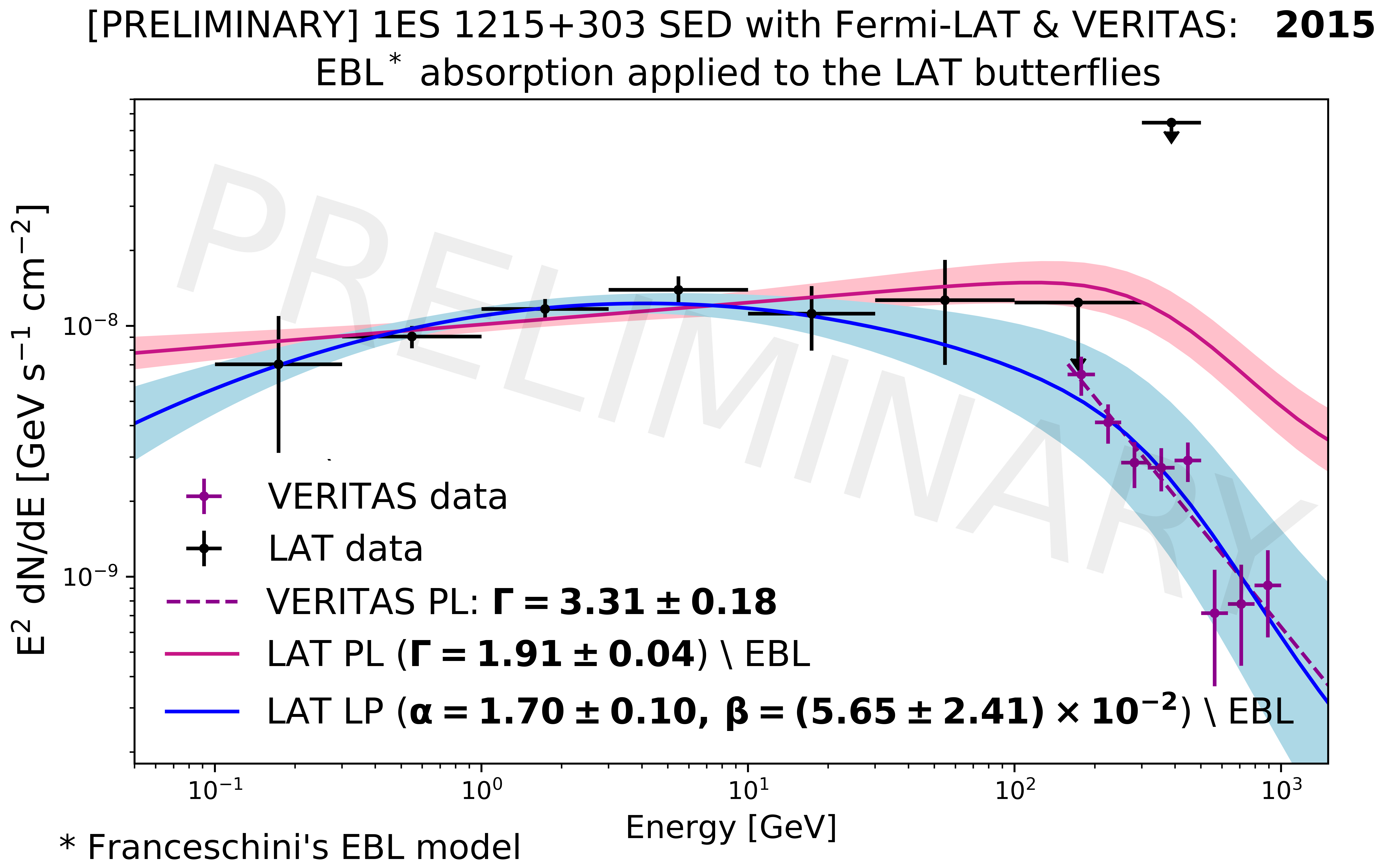}\includegraphics[width=0.49\textwidth]{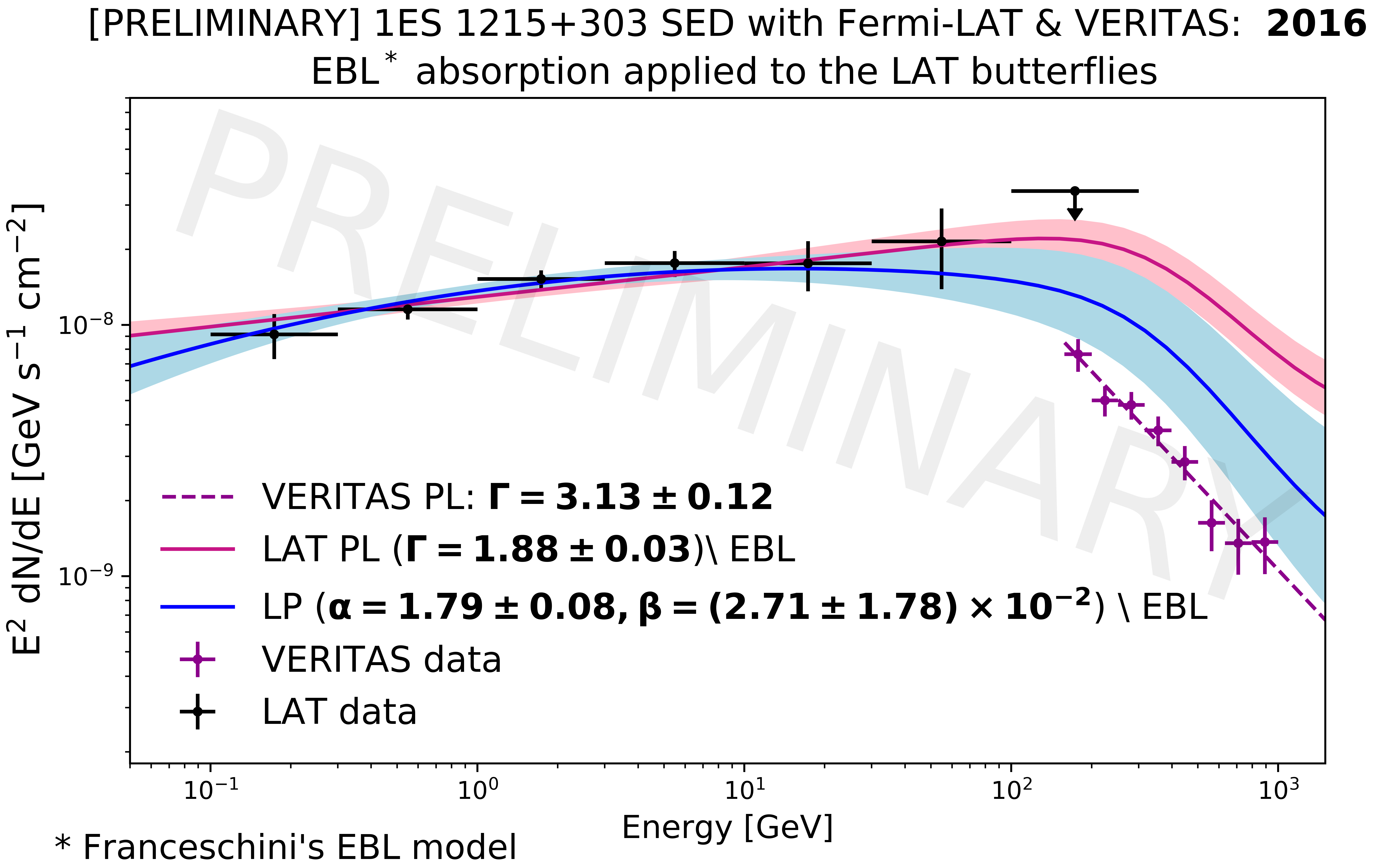}\\\vspace{1mm}
     \includegraphics[width=0.99\textwidth]{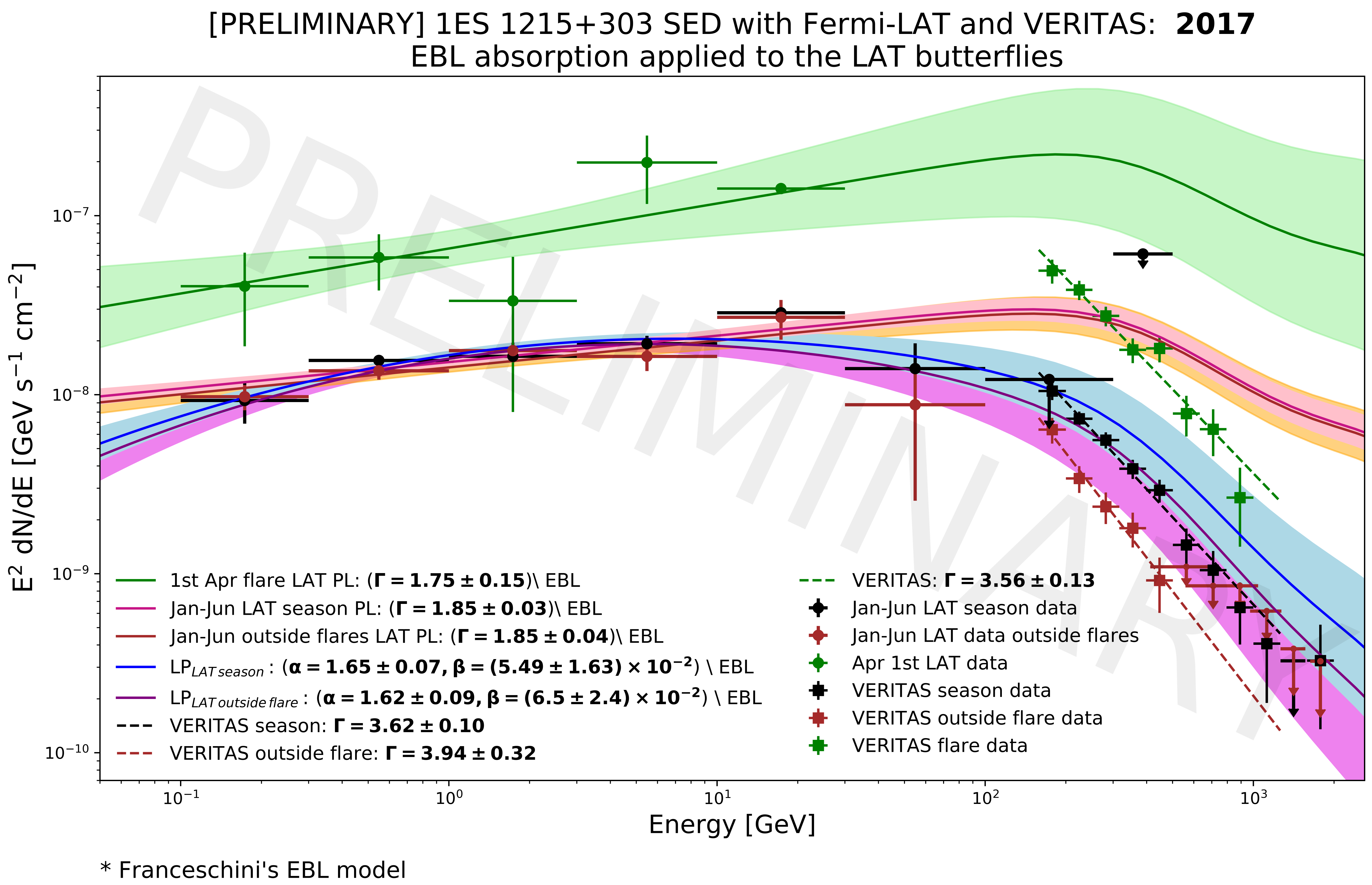}
\caption{SEDs of years 2015 (top left), 2016 (top right) and 2017 (bottom). The flaring and quiescent states SEDs of year 2017 are additionally shown.}
      \label{fig:sed17}      
   \end{minipage} \end{figure}

Fig. \ref{fig:ltlc} shows the long-term light curves with VERITAS 1-day data points on the top and 3-day LAT data on the bottom. Zoomed-in image of 1-day light 
      curves are shown for the brightest flares of 1ES 1215+303 for the LAT data, and 5 min bins for the brightest VERITAS flare. The LAT upper limits are 
      shown as green downward triangles. The VERITAS light 
      curve includes archival data up to 2014; the 2015-2017 data were not previously published. An additional interesting feature can be observed in Fig. \ref{fig:1030}, 
where a possible increase in the average flux of 1ES 1215+303 is shown in 30-day and 10-day-binned light curves. 
A detailed analysis of the quiescent state will be necessary in order to understand this possible trend. Moreover, we report the detection of a
      coincidental GeV-TeV flare on 2017-04-01. The SED of this flare can be seen in Fig. \ref{fig:sed17} together with the SED of the 
      entire 2017-01-01 to 2017-06-30 season. The SED of the season without the two LAT-detected flaring epochs is also shown. 
      In Fig. \ref{fig:sed17}, the GeV-TeV data are observed to be well-connected in all of the SEDs. The EBL-absorbed \cite{fran} LAT spectral models were extrapolated to the VERITAS energy range.
We observe that the LP SED is more compatible than the PL model with the TeV observations.
\vspace{-2mm}
     \vspace{-7mm}   
\section{Discussion}
The LAT data have demonstrated a preference for a log-parabola fit to the GeV spectrum (a power-law fit was rejected at the 5.26$\sigma$ level with the first 8 years of LAT data).
This indicates that the $\gamma$-ray spectrum of this blazar begins to turn over in the LAT energy range, i.e., at energies below those that can be attributed purely to absorption 
by the EBL thus indicating curvature intrinsic to the source itself. The EBL-absorbed LAT log-parabola spectrum is consistent with the spectrum measured by VERITAS at higher energies, 
and therefore already subjected to the effects of EBL absorption. Unlike what has been observed for other blazars at TeV \cite{djan}\cite{aliu} and GeV energies \cite{roop}, 
the spectral index of this blazar does not show significant evidence for hardening when it enters a higher flux state (Fig. \ref{fig:sed17}). 
This relationship will be investigated further in our ongoing more-detailed analysis \cite{ferver} as this relationship has been shown to be complex when shorter-term spectra were derived - e.g. for 
the case of Mrk 421 \cite{abey}. \\
The fact that we were able to measure the GeV flux of this blazar in such unprecedented detail over a ten-year span has allowed us to evaluate the evolution of the flux with time. 
An interesting feature that we have already found is an apparent increasing trend in its flux over the duration of the observations (Fig. \ref{fig:1030}). We find that, for the 
GeV $\gamma$-ray flux, an increasing-flux model is preferred to a constant flux at more than 15$\sigma$ level. Additionally, a preliminary yearly study of the complete LAT data set
has shown some indication of a spectral-hardening trend of 1ES 1215+303 as a function of time. We are investigating this potential trend in the flux and spectral index, 
excluding the flaring epochs, and will present a more in-depth study in a subsequent publication that is in preparation between the {\it Fermi}-LAT and VERITAS Collaborations \cite{ferver}.\\\newline
\begin{minipage}{0.43\textwidth}
In conclusion, this unique data set will help us to advance our understanding of the underlying physical processes in blazar jets through the investigation of the variability,
the search for periodicity and the cross-correlation studies between different wavelengths. Moreover, further investigation of the trend of long-term increasing flux and 
potential spectral hardening is ongoing.
\end{minipage}
\hfill
\begin{minipage}{0.54\textwidth}
\textbf{Table 1:} Summary of preliminary results\footnote{Refer to the paper in preparation for the final numbers.}.\\
{\tiny
\begin{tabular}{c c c c c}
\hline\hline \\ 
\textbf{Instrument} & \textbf{Dates} & \textbf{Live time} [min.] & \textbf{Sign.} & $\Gamma$\\ [0.5ex]
\hline\hline\\ 
 & 2015-01-17 to 2015-05-19 & 979.4 & 17.1$\sigma$ & -3.31$\pm$0.18\\
VERITAS & 2016-01-12 to 2016-05-09 & 1324 & 26.4$\sigma$ & -3.13$\pm$0.12 \\
 & 2017-01-03 to 2017-05-29 & 1473.95 & 33.0$\sigma$ & -3.62$\pm$0.10 \\
 & 2015-01-17 & 114.3 & 17.1$\sigma$ & -2.96$\pm$0.18 \\
 & 2017-04-01 & 150.1 & 28.9$\sigma$ & -3.56$\pm$0.13 \\
 \hline\\
 & 2015-01-01 to 2015-06-30 &  & 37.4$\sigma$ & -1.91$\pm$0.04 \\
{\it Fermi}-LAT & 2016-01-01 to 2015-06-30 &  & 44.4$\sigma$ & -1.88$\pm$0.03 \\
 & 2017-01-01 to 2017-06-30 &  & 54.7$\sigma$ & -1.85$\pm$0.02 \\
 & 2008-08-04 to 2017-09-04 &  & 139$\sigma$ & -1.92$\pm$0.01 \\
 & 2017-04-01 &  & 11.2$\sigma$ & -1.75$\pm$0.15 \\
\hline\hline
\end{tabular}}
\end{minipage}\vspace{4mm}
\begin{minipage}{1\textwidth}
See \url{https://veritas.sao.arizona.edu} for the VERITAS acknowledgements.\\
The {\it Fermi}-LAT Collaboration acknowledges support for LAT development, operation and data analysis from NASA and DOE (United States), CEA/Irfu and IN2P3/CNRS (France), ASI and INFN (Italy),
MEXT, KEK, and JAXA (Japan), and the K.A.~Wallenberg Foundation, the Swedish Research Council and the National Space Board (Sweden). Science analysis support in the operations phase from INAF (Italy)
and CNES (France) is also gratefully acknowledged. This work performed in part under DOE Contract DE-AC02-76SF00515.\end{minipage}

\bibliographystyle{JHEP}

\end{document}